\documentclass[conference]{IEEEtran}
\IEEEoverridecommandlockouts
\usepackage{cite}
\usepackage{amsmath,amssymb,amsfonts}
\usepackage{algorithmic}
\usepackage{graphicx}
\usepackage{textcomp}
\usepackage{xcolor}
\def\BibTeX{{\rm B\kern-.05em{\sc i\kern-.025em b}\kern-.08em
    T\kern-.1667em\lower.7ex\hbox{E}\kern-.125emX}}
\begin{document}

\title{Design and Manufacture of Flexible Epidermal NFC Device for Electrochemical Sensing of Sweat\\
\thanks{Work funded by Regione Lazio, project E-CROME (biosEnsori su Carta wiReless per la telemedicina in Oncologia e la misura di eMocromo ed Elettroliti; Development of NFC interface sensors for the measurement of biomarkers in blood), CUP: E85F21001040002.}
}

\author{\IEEEauthorblockN{Adina Bianca Barba}\IEEEauthorblockA{\textit{Pervasive Electromagnetics Lab} \\\textit{University of Rome Tor Vergata}\\Rome, Italy\\adinabianca.barba@alumni.uniroma2.eu}\and\IEEEauthorblockN{Giulio Maria Bianco}\IEEEauthorblockA{\textit{Pervasive Electromagnetics Lab} \\\textit{University of Rome Tor Vergata}\\Rome, Italy\\giulio.maria.bianco@uniroma2.it} \and \IEEEauthorblockN{Luca Fiore} \IEEEauthorblockA{\textit{NanoBioSensing Lab} \\ \textit{University of Rome Tor Vergata}\\ Rome, Italy \\ luca.fiore@uniroma2.it} \and \IEEEauthorblockN{Fabiana Arduini} \IEEEauthorblockA{\textit{NanoBioSensing Lab} \\ \textit{University of Rome Tor Vergata}\\ Rome, Italy \\ fabiana.arduini@uniroma2.it } \and \IEEEauthorblockN{Gaetano Marrocco} \IEEEauthorblockA{\textit{Pervasive Electromagnetics Lab} \\ \textit{University of Rome Tor Vergata}\\ Rome, Italy\\ gaetano.marrocco@uniroma2.it} \and \IEEEauthorblockN{Cecilia Occhiuzzi}\IEEEauthorblockA{\textit{Pervasive Electromagnetics Lab} \\ \textit{University of Rome Tor Vergata}\\ Rome, Italy\\ cecilia.occhiuzzi@uniroma2.it}} 
\maketitle

\begin{abstract}
Flexible and epidermal sensing devices are becoming vital to enable precision medicine and telemonitoring systems. The NFC (Near Field Communication) protocol is also becoming increasingly important for this application since it is embedded in most smartphones that can be used as pervasive and low-cost readers. Furthermore, the responder can be passive and can harvest enough power to perform electromagnetic sensing. Finally, the NFC coils are robust to bending and to the human body's presence. This contribution details the design of a new flexible device, including an electrochemical sensor communicating through the NFC protocol. A spiral NFC antenna is designed, and a manufactured prototype is experimentally tested to quantify the robustness to the inter-wearer variability and the bending. Lastly, the sensory data retrieval is validated by comparison with a portable potentiostat. The realized sensor can be comfortably worn and be easily read by smartphones independently from the wearer and from the point of application and could be used in future for estimating the user's psycho-physical health by analyzing the body's sweat.
\end{abstract}


\section{Introduction}
Smart wearable and epidermal sensors are currently being investigated for healthcare applications following the paradigm of precision medicine \cite{Yu21Flexible, Jeong19, An17}. As an example, many devices for sensing biosignals like skin temperature and coughing \cite{Bianco22Survey} or analytes in the sweat \cite{Jo21AReview} have been recently proposed. However, the wireless communication of the sensed data is still a major concern together with the use of local power supply, data security and the use of low-cost readers  for preserving user comfort and widespread diffusion \cite{Pillai21Advances, Mazzaracchio21Medium}. The Near Field Communication (NFC) protocol is an excellent option since the short reading distances can protect the privacy of the patient \cite{DiRienzo20Evaluation}. Moreover, most smartphones can read NFC devices that are nowadays a pervasive technology and will become even more utilized when interoperability issues will have been overcome \cite{Erb20}. The HF (high frequency) spiral antennas of NFC tags \cite{Chen20ASimple,DelRioRuiz17} are remarkably robust to bending \cite{Jiang17eTextile,Le20} and the environment \cite{Lazaro18ASurvey}, enabling integration in flexible sensors where limited space is available \cite{Iqbal14Wearable}. \par
\begin{figure}[tbp]
  \centering
  \begin{tabular}{cc}
  \includegraphics[width=40mm]{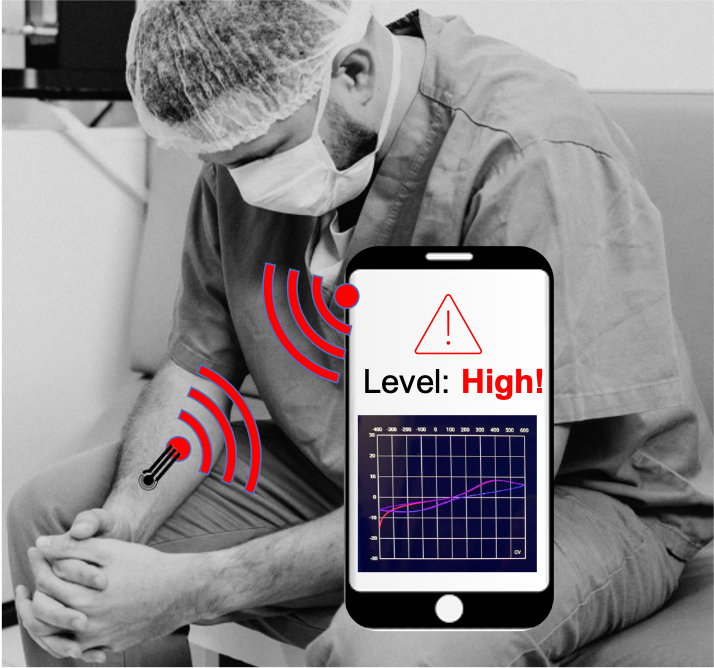} & \includegraphics[width=40mm]{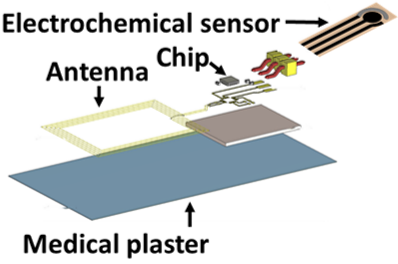}\\
(a) & (b)
\end{tabular}
  \caption{(a) Concept of sweat monitoring system comprising a passive, flexible and soft electrochemical sensor attached to the skin and capable of wirelessly communicating with a smartphone  (b) Numerical simulations of the designed NFC device. Exploded view of the model of the device.}
  \label{fig:ParametersExploded}
\end{figure}
The electrochemical sensors using commercial-off-the-shelves components communicating through NFC were proven extremely promising, mainly thanks to the highly efficient energy harvesting capability, which allows even passive microchips to perform complex chemical measurements such as amperometry and voltammetry \cite{Krorakai21Smartphone,Xu20Characterization}. This contribution hence details the design and test of a new epidermal biosensor employing the NFC protocol   [Fig.~\ref{fig:ParametersExploded}(a)]. The device is capable of performing real-time analysis of sweat and particularly of cortisol. Indeed, the cortisol levels can be linked to high levels of stress \cite{cay2018effect} and could enable precision monitoring, especially during sports activity or for workers during a hectic activity like soldiers or medical staff in hospital wards during the COVID-19 pandemic \cite{Bianco22Survey}. The device is totally passive and can be interrogated by smartphones. \par
\section{Design of the Flexible NFC Responder} 
The wireless epidermal device is composed of a spiral antenna, an NFC microchip, and an electrochemical sensor [Fig.~\ref{fig:ParametersExploded}(b)]. The design followed three main requirements: minimized size and weight, flexibility and biocompatibility of materials, reduced electromagnetic interference during interrogation. Therefore, the device has a rectangular shape whose overall size is similar to common skin patches. The antenna is located on the left side, far from the IC (integrated circuit) and the sensing circuitry, to reduce eventual detrimental effects resulting from the magnetic interaction with the smartphone \cite{Nalbantoglu16}. The antenna is made of copper micro-wires (diameter approximately 100 $\mu$m) directly laid down on a medical-grade self-adhesive membrane (Tegaderm \cite{Tegaderm}).  IC, tuning, and sensing circuitry are placed on a flexible  PCB. Finally, the sensor is located on the right side and is provided with a plug\&play connector to enable multiple usages.
The NFC SIC $4341$ (by Silicon Craft Technology) is the selected IC that is compatible with electrochemical sensing and can be read by a smartphone.\par
NFC tag  antenna can be modelled as an inductance $L$ having loss resistance $R$, and a tuning capacitance $C_{tun}$, while the IC as a capacitance-resistance parallel ($C_{chip}$ and $R_{chip}$). The NFC tag must also have a quality factor $Q=R_{chip}\left[2\pi f_r L\right]^{-1}$ lower than a maximum value $Q_{max}$ to channel all the frequencies contained in the HF-RFID spectrum. For targets employing the ISO/IEC $14443$A protocol like the considered one, $Q_{max}=25$ is a commonly accepted value \cite{Lazaro19Recent}.
By neglecting the parasitic capacitances, the resonance frequency $f_r$ of the equivalent circuit can be evaluated as $f_r\simeq \left[2\pi\sqrt{L\left(C_{tun}+C_{chip}\right)}\right]^{-1}$ \cite{Lazaro18ASurvey}.
%
%\begin{equation}\label{eq:ResonantFreq}
%f_r\simeq\left[2\pi\sqrt{L\left(C_{tun}+C_{chip}\right)}\right]^{-1}.
%\end{equation}
%
Even if the NFC protocol employs the carrier frequency of $13.56$ MHz, for the design, a higher $f_r=15.7$ MHz is considered to account for the frequency shift caused by the proximity of the reader's antenna during communications \cite{Krorakai21Smartphone}. 
Since $C_{chip}=50$ pF for the SIC $4341$ IC, by imposing $C_{tun}=27$ pF the resulting inductance is $L=1.36$ $\mu$H. \par
%
 %This value accounts for the entire device, including the IC. The quality factor of an NFC responder can be approximated as \cite{Lazaro19Recent}
%
%\begin{equation}\label{eq:qualityfactor}
%Q=\frac{R_{chip}}{\omega L}
%\end{equation}
%
%being $\omega=2\pi f$. We estimated  $R_{chip}=2375$ $\Omega$ by measuring with the Tagformance Pro (HF kit; by Voyantic) \cite{TagformancePro} the quality factor of the FlexSense board (configuration type 2, by Silicon Craft Technology), including the SIC $4341$ IC. $L$ of the board is from \cite{Krorakai21Smartphone}. \par
%
The NFC tag was designed through electromagnetic simulations\footnote{By CST Microwave Studio $2018$, frequency domain.}. A rectangular coil made of five turns of a thin copper wire is chosen as the antenna's geometry. The IC, instead, is placed on a pad (size $20$~mm~$\times$~$15$~mm~$\times$~$0.8$~mm) made of Kapton (modeled as a lossless dielectric with relative permittivity $3.3$ \cite{Simpson97}), and it is soldered to the antenna and to a connector through $35$-$\mu$m-thick copper traces [Fig. \ref{fig:ParametersExploded}(b)]. To account for the presence of the human body, the tag is centered on a multi-layer phantom having a square section $20$~cm~$\times$~$20$~cm and composed of dry skin thick $1$ mm, fat thick $1$ cm, and muscle thick $4$ cm (electrical properties at $f=15.7$ MHz from \cite{IFAC97}). \par
\begin{figure}[tbp] %figura L e Q simulato 
  \centering
  \includegraphics[width=50mm]{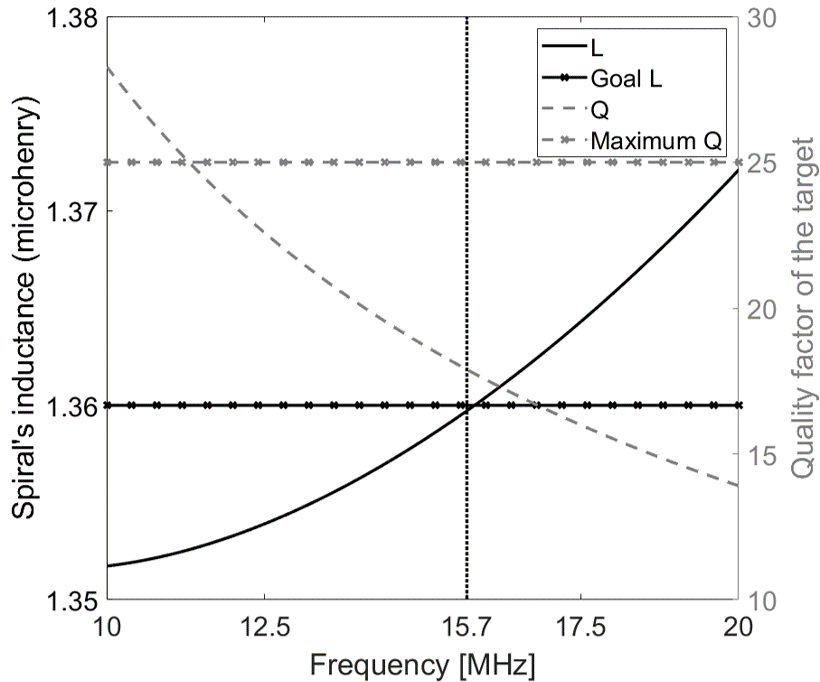}
  \caption{$L$ and $Q$ vs $f$ evaluated by numerical simulations. The design frequency of $15.7$ MHz is highlighted by a vertical line.}
  \label{fig:NumericalTarget}
\end{figure}
\begin{figure}[tb] 
\centering
\begin{tabular}{cc} 
% \includegraphics[height=40mm]{Images/StraghtSoleTag_2}  \\
%\includegraphics[height=50mm]{Images/prototypeBending.png}
 %\\ (a) & (b)  \\
 \\\includegraphics[height=48mm]{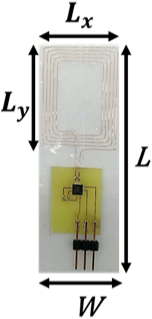} &  \includegraphics[height=48mm]{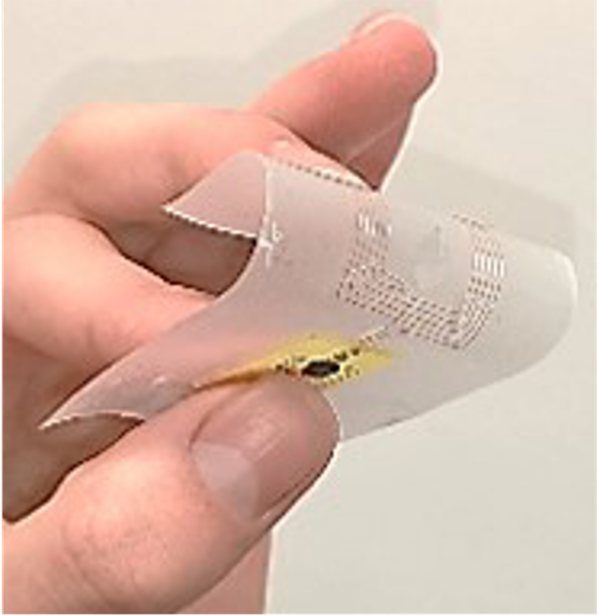} \\
 (a) & (b)  \\
\end{tabular}  
  \caption{(a) Prototype of the designed flexible device. Parameters  $W=31.5$ mm,  $L=65$ mm, $L_x=30$ mm, $s=0.9$ mm $L_y=21.7$ mm,  $w=40$ $\mu$m (b) Example of the flexed device.}
  \label{fig:InitiatorTarget}
\end{figure}

To achieve the needed inductance, four parameters were varied: the rectangle's longer side $L_x$, the rectangle's shorter side $L_y$, the space between the wires $s$, and the radius of the wires $w$. The parameters' values obtained by numerical simulations are resumed in the caption of Fig.~\ref{fig:InitiatorTarget}(a), and they return, at $f=15.7$ MHz, $L=1.36$ $\mu$H and $Q=18<25$ (Fig.~\ref{fig:NumericalTarget}). The total size $L\times W$ of the device allows for multiple positioning over the body while preserving the overall conformability and wearer comfort.\par
%
%\begin{figure}[tb] %figura L e Q simulato \centering\includegraphics[width=60mm]{Images/CounturA}\\(a) \\\includegraphics[width=60mm]{Images/CounturB}\\(b) \\  \includegraphics[width=60mm]{Images/CounturC}\\(c) \\\caption{Parametric charts of $L\left(f=15.7\textnormal{ MHz}\right)$ when varying (a) $N$ and $s$ fixed $\left(M,w\right)$, (b) $M$ and $N$ fixed $\left(s,w\right)$, and (c) $s$ and $W$ fixed $\left(M,N\right)$.}\label{fig:Counturs}\end{figure}
%
%\begin{table}[tbp] 
%  \centering
 % \caption{Values of the geometrical parameters of the spiral antenna.}  
%\begin{tabular}{|l|l||l|l|}
% \hline
% \bf{Parameter} & \bf{Value} & \bf{Parameter} & \bf{Value} \\
% \hline\hline
 % $W$ & $31.5$ mm & $L$ & $65$ mm\\
%\hline
%  $L_x$ & $30$ mm & $s$ & $0.9$ mm \\ 
% \hline
%  $L_y$ & $21.7$ mm & $w$ & $40$ $\mu$m \\
%  \hline
%\end{tabular}  \label{tab:ParametersValues}
%\end{table}
%

\section{Manufacturing and Measurements}
\subsection{Manufacturing}
For the sake of prototyping, the copper traces are obtained through a milling machine on an FR-$4$ board instead of the Kapton \cite{Mipec}. The spiral coils made of thin, flexible and moderately stretchable \cite{Miozzi20Radio} copper wire are placed by machines on the Tegaderm 3M medical plaster \cite{Mecstar16}. Then, the coil and the traces are manually soldered together afterwards. The overall epidermal biosensor is entirely deployed on the transparent and breathable medical plaster [Fig.~\ref{fig:InitiatorTarget}(a)]. Even though the FR-$4$ is rigid, the overall sensor is highly flexible for most of the space being occupied by the spiral made of wires, and it can be comfortably worn all over the body [Fig.~\ref{fig:InitiatorTarget}(b)].\par

\subsection{Communication test}
\begin{figure}[tb] %figura reading range aria e sui corpi con deviazione standard 
  \centering
  \includegraphics[width=60mm]{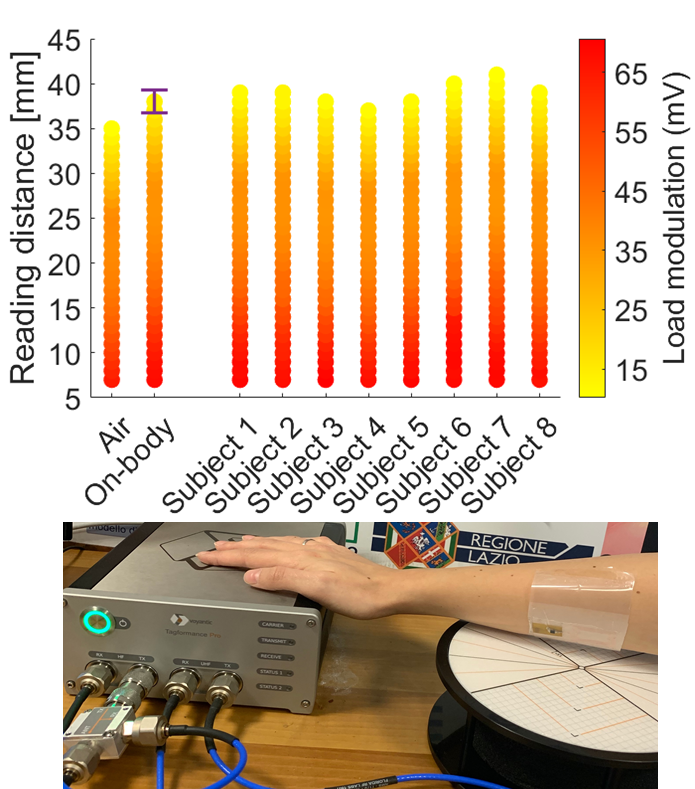} 
  \caption{Reading distances and load modulation when the biosensor is placed in air and when it is worn by eight different subjects. The "on-body" column shows the average over the eight wearers. Measurement setup at the bottom.}  \label{fig:InterBody}
\end{figure}

\begin{figure}[tb] %figura reading range aria e bending
  \centering
  \includegraphics[width=60mm]{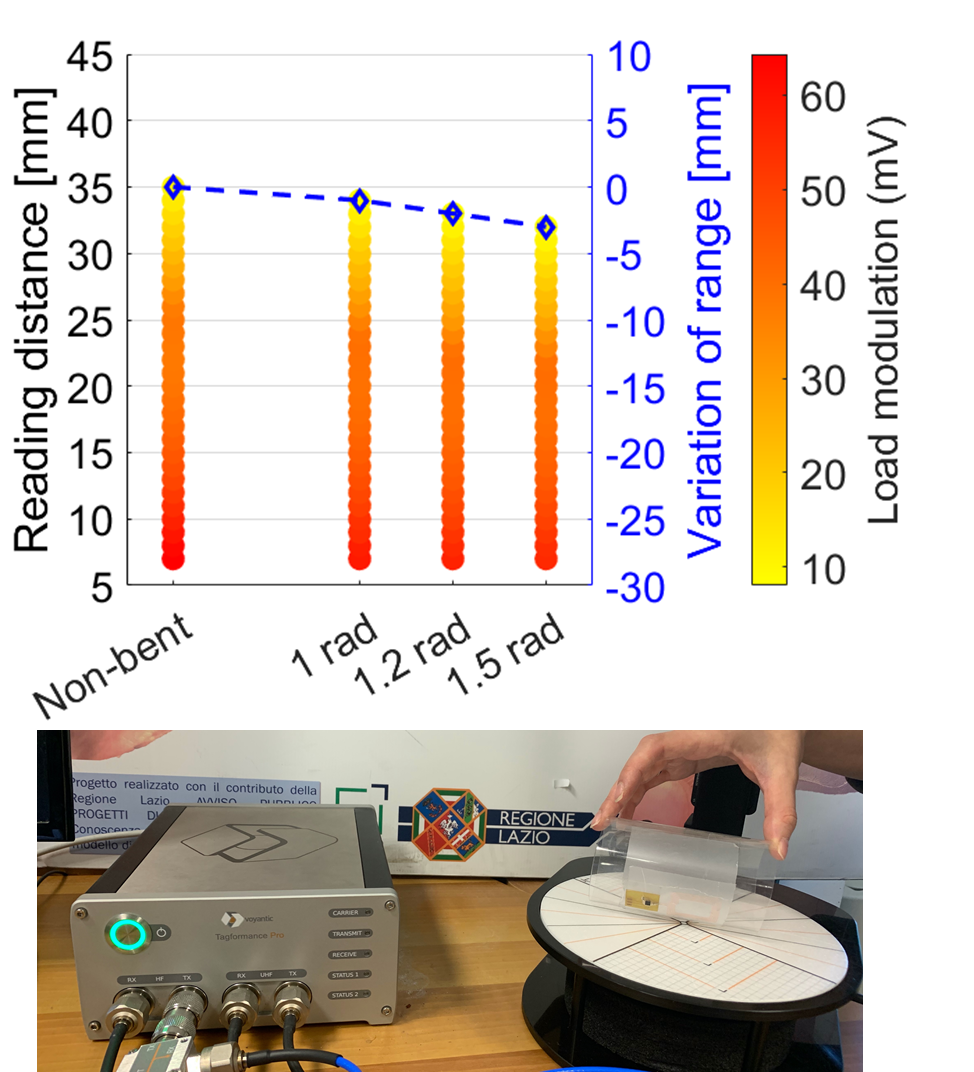}
  \caption{Load modulation, reading distances, and the corresponding reduction of communication range when the non-worn tag is bent for different bending angles. Measurement setup at the bottom.}
  \label{fig:Bending}
\end{figure}

The prototype is tested with the Tagformance Pro (C$60$ antenna) both in the air reference condition (namely, when the human body is absent) and when it is placed on the arm of eight different wearers (six females, two males; mean body mass index $22.7\pm 4.2$). The reading ranges and the quality of the magnetic coupling, which is expressed by the passive load modulation \cite{Paret16}, are depicted in Fig.~\ref{fig:InterBody}. The device can be read up to a distance of $40$ mm. The target responds slightly better on the human body than in the air ($+10\%$ reading range) since it was designed on a body phantom. The inter-subject variability \cite{Bianco20Experimentation} yields a standard deviation as low as $1.2$ mm. Similarly, the prototype is rather robust to the bending, and the maximum communication range is reduced by just $3$ mm when a bend angle of $1.5$ rad ($\sim85^o$ degrees) is applied by attaching the sensor to an empty plastic cylinder having a known radius (Fig.~\ref{fig:Bending}). Accordingly, a typical NFC range of about $3$ cm is expected independently from the user and the point of application of the sensor. In all the experiments, the coupling's quality follows the range variations, being remarkably stable and depending primarily on the reader-tag distance. \par
\subsection{Sensing test}
The capability of the NFC microchip of retrieving the data from the given sensor was finally assessed \cite{Krorakai21Smartphone}. The considered sensor of cortisol was developed starting from the chemical probe in \cite{Nappi21APlug, Mazzaracchio21Medium}, and, if wet, it returns an electrical current inversely proportional to the cortisol concentration. The current can be measured through chronoamperometry, i.e. by imposing a transient voltage and consequently measuring the current during a well-specified time interval.\par
For comparison, the sensor's output current is also measured with a portable potentiostat (Emstat Blue, by PalmSens). The NFC sensor is instead interrogated by an Android smartphone through the Chemister app by Silicon Craft.\par
\begin{figure}[tbp]
  \centering
  \includegraphics[width=60mm]{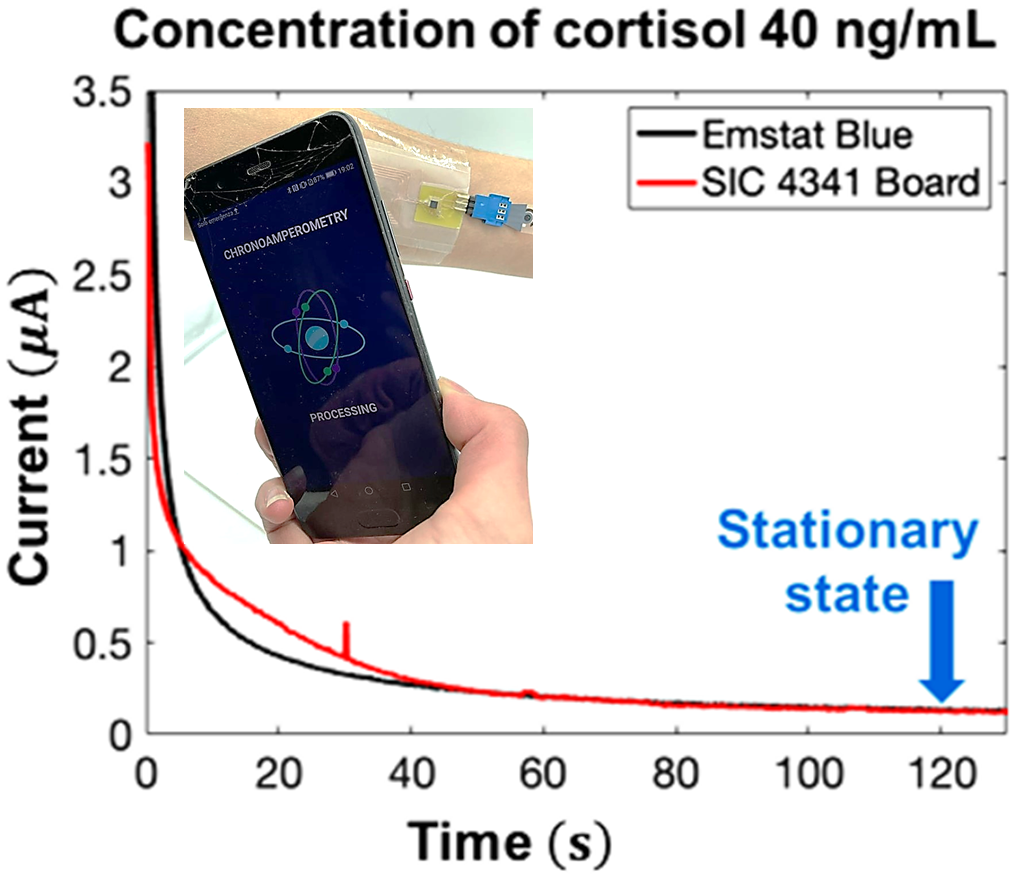}
  \caption{Electric current vs time curve when the biosensor is soaked by a cortisol concentration of $40$ ng/mL. In the inset: concept of the biosensory data retrieval by the Chemister app.}
  \label{fig:Calibrazione}
\end{figure}
The current vs time curves of the sensor produced by a concentration of cortisol of $40$ ng/mL are shown in Fig.~\ref{fig:Calibrazione}. The two curves are in close agreement and return the same current at their stationary states ($120$ s). Thus, the selected IC is viable for electrochemical sensing through the designed flexible device and smartphones.
\section{Conclusion}
A flexible epidermal sensing device communicating through the NFC protocol was designed and tested for application to cortisol sensing in sweat. The prototype is highly robust to the inter-user variability and bending thanks to the HF communications so that it achieved a reading range of about $3.5/4$ cm for every considered user and point of application on the body. The preliminary sensing test validates the accuracy of the retrieved sensory data, which is comparable with that of much more expensive bench equipment.
A miniaturized and fully flexible prototype is currently under development and will be presented during the Conference together with a complete test campaign in real conditions aimed at validating the sensing capabilities of the device over multiple subjects.
\bibliographystyle{IEEEtran}
\bibliography{Adina_ECROME_12} % file mwe.bib

\end{document}